\documentclass{emulateapj}

\usepackage{amsmath}
\usepackage{amssymb}
\usepackage{graphicx}
\usepackage{color}
\usepackage{natbib}
\usepackage{epsfig}
\usepackage{url}

\def\na{{New Astronomy}}                % New Astronomy
\def\rmxaa{{Revista Mexicana de Astronom{\' i}a y Astrof{\' i}sica}}

\def\gtaprx {\lower .1ex\hbox{\rlap{\raise .6ex\hbox{\hskip .3ex
	{\ifmmode{\scriptscriptstyle >}\else
		{$\scriptscriptstyle >$}\fi}}}
	\kern -.4ex{\ifmmode{\scriptscriptstyle \sim}\else
		{$\scriptscriptstyle\sim$}\fi}}}
\def\ltaprx {\lower .1ex\hbox{\rlap{\raise .6ex\hbox{\hskip .3ex
	{\ifmmode{\scriptscriptstyle <}\else
		{$\scriptscriptstyle <$}\fi}}}
	\kern -.4ex{\ifmmode{\scriptscriptstyle \sim}\else
		{$\scriptscriptstyle\sim$}\fi}}}

\newcommand{\pasa}{Publications of the Astronomical Society of Australia}

\begin{document}

\title{Balmer-Dominated Shocks Exclude Hot Progenitors for Many Type Ia Supernovae}
\author{ T.~E. Woods\altaffilmark{1}, P. Ghavamian\altaffilmark{2}, C. Badenes\altaffilmark{3}, M. Gilfanov\altaffilmark{4,5}}

\altaffiltext{1}{Monash Centre for Astrophysics, School of Physics and Astronomy, Monash 
University, VIC 3800, Australia}
\altaffiltext{2}{Department  of  Physics,  Astronomy  and  Geosciences,  Towson University, Towson, MD, 21252}
\altaffiltext{3}{Department of Physics and Astronomy and Pittsburgh Particle Physics, Astrophysics, and Cosmology Center (PITT-PACC), University of Pittsburgh,3941 O'Hara Street, Pittsburgh, PA 15260, USA}
\altaffiltext{4}{Max-Planck Institut f{\"u}r Astrophysik, Karl-Schwarzschild-Str. 1, D-85741 Garching, Germany}
\altaffiltext{5}{Space Research Institute, Profsoyuznaya 84/32, 117997 Moscow, Russia}
 
\begin{abstract}
The evolutionary mechanism underlying Type Ia supernova explosions remains unknown. Recent efforts to constrain progenitor models based on the influence that their high energy emission would have on the interstellar medium (ISM) of galaxies have proven successful. For individual remnants, Balmer-dominated shocks reveal the ionization state of hydrogen in the immediately surrounding gas. Here we report deep upper limits on the temperature and luminosity of the progenitors of four Type Ia remnants with associated Balmer filaments: SN 1006, 0509-67.5, 0519-69.0, and DEM L71. For SN 1006, existing observations of helium line emission in the diffuse emission ahead of the shock provide an additional constraint on the helium ionization state in the vicinity of the remnant.
Using the photoionization code Cloudy, we show that these constraints exclude any hot, luminous progenitor for SN 1006, including stably hydrogen or helium nuclear-burning white dwarfs, as well as any Chandrasekhar-mass white dwarf accreting matter at $\gtrsim 9.5\times10^{-8}M_{\odot}/$yr via a disk. For 0509-67.5, the Balmer emission alone rules out any such white dwarf accreting $\gtrsim 1.4\times10^{-8}M_{\odot}/$yr.
For 0519-69.0 and DEM L71, the inferred ambient ionization state of hydrogen is only weakly in tension with a recently hot, luminous progenitor, and cannot be distinguished from e.g., a relatively higher local Lyman continuum background, without additional line measurements.
Future deep
spectroscopic observations will resolve this ambiguity, and can either detect the influence of any luminous progenitor or rule out the same for all resolved SN Ia remnants.
\end{abstract}


\maketitle

\section{Introduction}

Type Ia supernovae (SNe Ia) are now understood to arise from the thermonuclear explosion of a carbon-oxygen white dwarf (WD) \citep{HKR13}. SNe Ia are most clearly identified by their spectra, characterised by strong absorption lines from ionized silicon and the conspicuous absence of hydrogen lines \citep{BW17}. These cataclysmic explosions play a critical role in the chemical evolution of the Universe and the energetics of the interstellar medium (ISM). Their calibration as standardizable candles revolutionized cosmology at the end of the last century. For a recent review of their observational features and evidence for their progenitors, see \cite{MMN14}.

Models for the evolution of SN Ia progenitors fall into one of two categories: accretion scenarios \citep[e.g.,][]{WI73,SB09}, and merger scenarios \citep[e.g.,][]{Webbink84,IS12}, often termed single- and double-degenerate channels. In the typical accretion scenario, a phase of either steady hydrogen \citep[e.g.,][]{Wolf13} or helium \citep[e.g.,][]{PTY14} shell-burning at the surface is invoked, which may last from tens of thousands to of order a million years. Steady nuclear-burning at the surface implies bolometric luminosities on the order of $10^{38}$ erg$\rm{s}^{-1}$ and effective temperatures in the range $10^{5}$ -- $10^{6}$ K \citep{vdHeuvel92}. Such objects are observed in the LMC, M31, and along those few lines of sight in the Galaxy where the column density is sufficiently low \citep{Greiner00}. In some cases, the product of a WD--WD merger may also persist for of order $10^{4}$ yrs with a similar temperature and luminosity to steady-burning WDs \citep[see e.g.,][]{Schwab16}. It is unclear, however, if these models correspond to objects which will explode as normal SNe Ia.

In order to test the viability of the accretion scenario, considerable effort has gone into attempting to detect the progenitors of contemporary, nearby SNe Ia in archival X-ray and optical data \citep[e.g.,][]{Li11, Nielsen13, Nielsen14, Kelly14}. However, this approach relies to an uncomfortable degree on luck; the community must wait for a sufficiently nearby supernova to occur where sufficiently deep pre-supernova imaging has taken place.

Recently, progress has been made in constraining the plausible contribution of any accretion scenario to the total observed rate. One approach is to consider the effect such a luminous, high temperature source population would have on the ISM of galaxies \citep[e.g.,][]{WG13,WG14}. In particular, accreting, nuclear-burning white dwarfs would generate luminous He II, [O I], and [O III] line emission \citep{Rappaport94, Remillard95}. In this way, hot luminous progenitors consistent with the standard accretion model were shown to contribute no more than a few percent of the SN Ia rate at late delay times \citep{Johansson14, Johansson16}.

Extending this method to individual objects in nearby, star-forming galaxies is complicated by the inability to average over properties of the ISM; the environment surrounding each SN Ia must be carefully considered. In an illustrative example, \cite{Graur14} placed an upper limit on the He II 4686\AA\ flux in a pre-supernova image of the vicinity of SN 2011fe, constraining the surface brightness of any putative ionized nebula. However, without an estimate of the density of the surrounding ambient ISM, the degree to which this can constrain the luminosity and temperature of the progenitor remains ambiguous.

The expanding shocks of supernova remnants serve as excellent probes of the density of the surrounding gas \citep[see e.g.,][]{Badenes07,Yamaguchi14}. At the same time, the optical emission at the shock front of many SN Ia remnants is observed to be dominated by both broad and narrow Balmer-line emission, resulting from the interaction of the expanding remnant with surrounding neutral hydrogen \citep[see e.g.,][for a review]{Heng10}. Modelling of the diffuse emission in a photoionized precursor ahead of the shock can allow one to infer the neutral fraction in the pre-shock gas. 
Taken together, these results can provide a robust measurement of the density and ionization state of the ISM surrounding many observed remnants. This constrains the nature of any ``relic'' nebula ionized by the progenitor \citep{Ghavamian03,Vink12}. 
This approach has recently been successful in excluding any otherwise-viable accretion scenario as the progenitor of Tycho's supernova \citep{Woods17}.

The efficacy of this approach, however, is not uniform for all remnants, depending in particular on the local ionizing background and the structure of the ambient ISM. In the following, we assess the viability of any hot, luminous progenitor scenario for four SN Ia remnants with observed Balmer-dominated shocks, where the density and ionization state of the surrounding ISM has been measured or strongly constrained: the Galactic remnant SN 1006, as well as DEM L71, 0509-67.5, and 0519-69.0 in the Large Magellanic Cloud. In section \ref{environment}, we discuss the physics of relic nebulae, and the ionization state of the ISM surrounding the chosen remnants as revealed by the observed hydrogen and, for SN 1006, helium line emission. The importance of this additional constraint on the environment of SN 1006 is made clear in section \ref{nebulae}, wherein we assess the viability of the accretion scenario for each remnant by comparison with models for putative relic nebulae ionized by SN Ia progenitors.
Finally, we discuss further prospects for reconstructing the progenitor characteristics of recent and historical supernovae with ongoing and future deep spectroscopic studies in section \ref{conclusion}.

\section{SN Ia remnants and their surrounding ISM}
\label{environment}
\subsection{Lifetime and properties of relic nebulae}

Most SNe Ia do not explode within the particularly dense regions of the ISM associated with recent star formation, as is the case for core-collapse supernovae with much shorter delay times. Rather, the typical ambient medium is expected to be charateristic of the warm ionized and neutral phases of the ISM (0.1 -- 1$\rm{cm}^{-3}$, T~$\approx 10^{4}K$). This is consistent with numerical shock models and X-ray observations of known remnants \citep[e.g.,][]{Yamaguchi14}. If the progenitors of SNe Ia are sufficiently hot and luminous (i.e., T $\gtrsim$ $5\times 10^{4}K$, L $\sim$ $10^{38}$erg$\rm{s}^{-1}$), they will significantly ionize their surrounding ISM \citep{Rappaport94,WG16} out to a characteristic Str{\" o}mgren radius:

\begin{equation}
\rm{R}_{\rm{S}} \approx 35\rm{pc}\left(\frac{\dot N_{\rm{ph}}}{10^{48}\rm{s}^{-1}}\right)^{\frac{1}{3}}\left(\frac{\rm{n}_{\rm{0}}}{1\rm{cm}^{-3}}\right)^{-\frac{2}{3}}\label{RS}
\end{equation}

\noindent where $\dot N_{\rm{ph}}$ is the ionizing photon luminosity of the progenitor, and $\rm{n}_{\rm{0}}$ is the density of the surrounding ISM. For variable sources \citep{CR96}, it is the time-averaged ionizing photon luminosity that is the quantity of interest \citep[see discussion in][]{Woods17}. In practice, this simple picture is broken for higher temperature sources (T $\gtrsim$ few $\times$ $10^{5}K$), as higher energy photons penetrate deeper into the neutral gas, significantly broadening the Str{\" o}mgren boundary that marks the transition between ionized and neutral media \citep[e.g.,][]{WG16}. For this reason, measurement of (or upper limits on) additional emission lines characteristic of warm, partially-ionized regions (e.g., [O I] $\lambda$6300]) would be invaluable in differentiating between a very hot progenitor and a high Lyman continuum background (i.e., if the H ionization fraction is relatively high in the vicinity of a SN Ia remnant).

After the supernova explosion, emission from the source ceases; however, a ``relic'' nebula will persist for the recombination timescale of the ISM:

\begin{equation}
\tau _{\rm{rec}} = 1/(\alpha _{\rm{B}}(H^{0},T)\rm{n}_{\rm{e}}) \approx 10^{5} \left(\frac{\rm{n}_{\rm{e}}}{1\rm{cm}^{-3}}\right)^{-1} \rm{years} \label{timescale}
\end{equation}

\noindent where $\alpha _{\rm{B}}$ is the recombination coefficient for neutral hydrogen ($H^{0}$) at a given gas temperature $T$, and $n_{\rm{e}}$ is the electron density. This is comparable to the typical lifetimes of supernova remnants themselves \citep[see e.g.,][]{MB10}. Therefore, if a SN Ia remnant is observed to be interacting with neutral gas, this can provide a strong constraint on the size of the Str{\" o}mgren region (and thus the temperature and luminosity) of the progenitor for more than 100,000 years prior to explosion.

\subsection{Type Ia Supernovae with Balmer-Dominated Shocks}

Many remnants of known or suspected SNe Ia exhibit regions of detectable broad and narrow Balmer line emission along the forward shock. Such ``Balmer-dominated'' shocks arise when the forward shock (v $\gtrsim$ 1000km$\rm{s}^{-1}$) overruns neutral interstellar gas \citep{CR78}. A fraction of the cold neutral hydrogen atoms entering the shock will be collisionally-excited prior to being fully ionized either by collision or charge transfer. The ensuing radiative decay produces the narrow component of the observed Balmer emission, whose width is set by the pre-shock temperature of the ISM. The broad component originates from charge exchange between cold ambient neutrals overrun by the shock and hot protons behind the shock \citep[for futher details, see e.g.,][]{CKR80}.

\begin{table}[ht]
\begin{center}
\caption{SN remnants considered, and constraints on their surrounding environment}\label{data}
\begin{tabular}{l c c c c c c}
Remnant & r  & $\rm{f}_{\rm{H^{0}}}$ & $\rm{f}_{\rm{He^{0}}}$ & $\rm{n}_{0}$\\
\hline
 & pc & & & $\rm{cm}^{-3}$\\
\hline
SN 1006 & 10 & $\gtrsim$0.1 & $\gtrsim$ 0.7 & $<$0.4\\
0509-67.5 & 4 & $>$0.4 & unknown & $<$0.3\\
0519-69.0 & 4 & 0.4--0.5  & unknown & $<$2.4\\
DEM L71 (West) & 9.0 & 0.2--0.4 & unknown & 0.5\\
DEM L71 (East) & 6.8 & 0.2--0.4 & unknown & 1.5\\
\hline
\end{tabular}
\end{center}
\end{table}

The ratio of broad to narrow Balmer emission ($I_{\rm{B}}/I_{\rm{N}}$) in these shocks depends critically on the shock velocity ($\rm{V}_{\rm{shk}}$), the temperature equilibration between ions and electrons in the post-shock gas ($\rm{T}_{\rm{e}}/\rm{T}_{\rm{P}}$), and the hydrogen neutral fraction ($\rm{f}_{\rm{H^{0}}}$) of the pre-shock medium \citep{Ghavamian01, Ghavamian02, Ghavamian03}. Of these, the shock velocity can be constrained by the width of the broad component, as well as X-ray observations and numerical models of the advancing shock. The temperature equilibration is, unfortunately, a free parameter in modelling $I_{\rm{B}}/I_{\rm{N}}$, and degenerate with the hydrogen neutral fraction \citep{Ghavamian01}. The broad-to-narrow ratio also nominally depends on the efficiency of cosmic ray acceleration \citep{Morlino12,Morlino13}. Nonetheless, numerical modeling of $I_{\rm{B}}/I_{\rm{N}}$ can provide us with a minimum plausible hydrogen neutral fraction.

In the following, we provide an outline of existing measurements of the size of four remnants, as well as the density and ionization state of their surrounding ambient medium, summarized in table \ref{data}: in our Galaxy, SN 1006, and in the LMC, DEM L71 (0505-67.9), 0509-67.5, and 0519-69.0. All remnants are unambiguously identified as being Type Ia based on their X-ray spectra, and particularly the high abundance of iron and its ionization state in their ejecta \citep{Yamaguchi14,Maggi16,PB17}. The natures of 0509-67.5 and 0519-69.0 have also been independently confirmed by light echo spectroscopy \citep{Rest05}. See \cite{PB17} for further discussion.

\subsubsection{SN 1006}

\cite{Ghavamian02} studied the optical emission in the northwest rim of the remnant of SN 1006. They obtained a deep long-slit spectrum of this filament, detecting He I 6678\AA\ emission in addition to H$\alpha$, H$\beta$, and H$\gamma$, as well as a marginal (1.5 $\sigma$) measurement of He II 4686\AA. Modeling the shock emission, they found that one could only reproduce the observed $I_{\rm{B}}/I_{\rm{N}}$ ratio given a pre-shock neutral hydrogen fraction of $\rm{f}_{\rm{H}^{0}} \gtrsim 0.1$. At the same time, fitting the He I/He II and He I/H$\alpha$ ratios implied a pre-shock neutral helium fraction ($\rm{He}^{0}/\rm{He}$) of $\rm{f}_{\rm{He}^{0}} \gtrsim$70\%. These models also indicated a low electron-proton equilibration at the shock front ($T_{e}/T_{p} \leq 0.07$) and a high shock velocity (2890$\pm$100km$\rm{s}^{-1}$). Finally, \cite{Ghavamian02} argued the high ionization state of hydrogen and low ionization state of helium in the environment of SN 1006 are plausibly consistent with photoionization by the background Galactic Lyman continuum. In particular, they exclude significant ionization by emission from the reverse shock or an extreme UV flash from the original supernova itself.

Measurements of the distance to SN 1006 have converged on $\sim$2kpc, yielding a remnant radius of $\approx$10pc \citep[see discussion in][]{Ghavamian02,Raymond07}. Modelling of the X-ray emission suggests the Fe K$\alpha$ luminosity of the shocked gas and its centroid energy are consistent with a pre-shock gas density of $n_{0}$ $<$ $1\rm{cm}^{-3}$ \citep{Yamaguchi14}. Earlier estimates based on the global X-ray emission suggests pre-shock densities as low as 0.05 -- 0.1 $\rm{cm}^{-3}$ may be plausible \citep{Hamilton86}. \cite{Raymond07} compared numerical models of the H$\alpha$ emissivity with a deep {\it Hubble Space Telescope} ({\it HST}) image of SN 1006 to derive a particle density of $0.25 \rm{cm}^{-3} \leq n_{0} \leq 0.4 \rm{cm}^{-3}$ for the northwestern, Balmer-emitting quadrant of the remnant. In our subsequent analysis, we take a mean ISM density of $0.4 \rm{cm}^{-3}$ as a conservative upper bound; for lower densities, our results in section \ref{results} becomes more constraining, with the Str{\" o}mgren radius scaling as $\propto n_{0}^{-2/3}$ (c.f. eq. \ref{RS}). Note that the ionization fractions derived in \cite{Ghavamian02} are insensitive to the total pre-shock gas density $n_{0}$.

\subsubsection{Magellanic Supernova Remnants} 

\cite{Ghavamian03} study the entire remnant of DEM L71 using Fabry-Perot imaging spectroscopy, with pre-shock densities inferred from modeling {\it Chandra} ACIS-S spectra \citep{RGH03}. The latter find a range of pre-shock densities along different regions of the rim of the remnant, roughly bounded by the range $\approx 0.5$--$1.5\rm{cm}^{-3}$. In particular, $n_{0} \approx 0.5\rm{cm}^{-3}$ is consistent with the majority of the eastern rim of the remnant, while in portions of the north-west rim $n_{0} \approx 1.5\rm{cm}^{-3}$. Modeling the observed collisionally-excited hydrogen-Balmer emission, \cite{Ghavamian03} conclude that a wide range in neutral fraction may be plausible, but that $\rm{f}_{\rm{H}^{0}}\gtrsim$ 0.1 would be necessary in order to produce detectable H$\alpha$ emission. 

\cite{Ghavamian07} revisited DEM L71 in the far-UV, as well as the LMC remnants 0509-67.5, 0519-69.0, and 0548-70.4, using spectra obtained from the {\it Far Ultraviolet Spectroscopic Explorer} ({\it FUSE}). The ratio of fluxes in Ly$\beta$ and O {\sc VI} 1032\AA\ depends linearly on the pre-shock neutral fraction; the observed ratio in DEM L71 requires a hydrogen neutral fraction of 20--40\%, consistent with previous limits. 

No UV lines were found for the remnant 0548-70.4. For the remnants 0509-67.5 and 0519-69.0, the same method constrains the neutral hydrogen fraction in the surrounding ISM to $>$40\%. The best-fitting ambient densities can be found from comparing the present size and ionization state of the remnant with hydrodynamical models: recently, \cite{Kosenko14} found $n_{0} = 0.1$--$0.3\rm{cm}^{-3}$ for 0509-67.5, and $n_{0} = 1.0$--$2.0\rm{cm}^{-3}$ for 0519-69.0, although \cite{Kosenko10} report densities as high as $2.4\rm{cm}^{-3}$. As a conservative estimate, we adopt 0.3$\rm{cm}^{-3}$ and 2.4$\rm{cm}^{-3}$ as our fiducial ISM densities for 0509-67.5 and 0519-69.0 respectively. The radius of the forward shock for both remnants is $\sim$4pc, for a LMC distance of 50kpc.

\section{Constraints on the nature of the progenitors of nearby SNe Ia}
\label{nebulae}

\subsection{Flash Ionization}

\begin{figure*}[ht]
\hskip-0.75cm \includegraphics[width=1.05\textwidth]{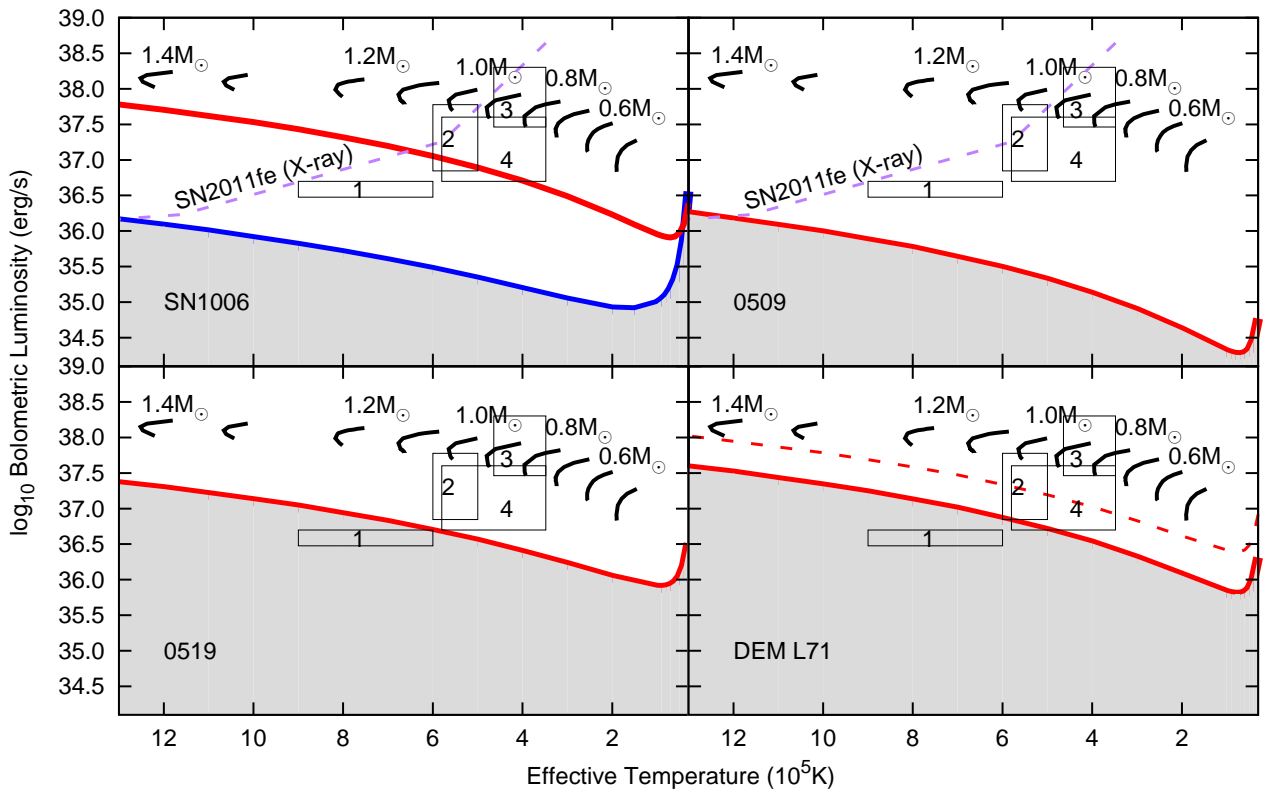}
\caption{Upper limits on the bolometric luminosity of the progenitors of SN 1006, 0509-67.5, 0519-69.0, and DEM L71 (0505-67.9). Red and blue lines denote upper limits given the inferred H I and He I neutral fractions, respectively. Also shown for comparison (in purple, dashed lines, for SN 1006 and 0509-67.5) is the most constraining upper limit to date from pre-supernova archival X-ray data, for SN 2011fe \citep{Nielsen12}. For DEM L71, red dashed (n~=~1.5$\rm{cm}^{-3}$, r = 6.8pc) and solid (n~=~0.5$\rm{cm}^{-3}$, r = 9pc) lines denote upper limits given the ionization state of hydrogen as measured at different points along the shock front.  Black lines denote the nuclear-burning accreting white dwarf models of \cite{Wolf13}. Note that for ease of reading, only every second model is labelled (0.51$M_{\odot}$ to 1.4$M_{\odot}$). Black boxes mark known persistent supersoft sources with well-established temperatures and luminosities: 1.~CAL 87; 2.~1E 0035.4-7230; 3.~RX J0513.9-6951; and 4.~CAL 83 \citep{Greiner00,Starrfield04}. Note that CAL 87 is viewed nearly edge-on, and its unobscured luminosity is likely much higher \citep{Ness13}.}\label{allSNe}
\end{figure*}

We may now investigate the extent to which the presence of neutral hydrogen at the present radii of all remnants considered above constrains the release of ionizing photons in the last $\tau _{\rm{rec}} \approx$ 100,000 years. \cite{CR78} were the first to consider the simplest case, a ``burst'' of ionizing photons from the supernova itself. For a present shock radius $R$ and mean ambient density $n$, the total number of hydrogen-ionizing photons emitted in such a burst must have been fewer than $4\pi R^{3}n/3 \approx (0.2-6)\times 10^{58}$ for SN 1006, 0509-67.5, 0519-69.0, and DEM L71 (c.f. table \ref{data}). Notably, this upper bound applies not only to the shock breakout from the supernova itself, but to any short-lived phase which the progenitor or its companion underwent immediately preceding or following the explosion, such as any briefly hot, luminous surviving white dwarf donors \citep{SS17}. Understanding the precise spectral appearance of such objects, however, awaits detailed radiative transfer calculations. Therefore, we leave any further consideration of the plausibility of this model for a future study \citep[see also][for a complementary direct search for surviving white dwarf donors in SN 1006]{Kerzendorf17}.

The simplified approach given above is inadequate for constraining the luminosity of any long-lived hot progenitor with an associated steady-state photoionized nebula, and ignores the typical path length of the ionizing photons, relevant for the broad partially-ionized zones associated with very high temperature sources. In the following, we assess the viability of accreting and steadily nuclear-burning white dwarf progenitor models using a detailed photoionization code.

\subsection{Numerical simulations with Cloudy}

In order to constrain the time-averaged photoionizing luminosity of a long-lived supernova progenitor, we directly compare the hydrogen and/or helium ionization fractions measured at the present forward shock radii of each Balmer-dominated supernova remnant with numerical photoionization models of the expected surrounding nebulae for a given source luminosity, source temperature, and surrounding ambient density. In this way, we can derive a maximum plausible source luminosity at any given progenitor temperature, above which the expected ionization fraction would be greater than that which is observed today at the present radius of the shock \citep{Woods17}.

Cloudy \citep[v13.03,][]{Ferland13} is an open-source software package which computes the conditions in an arbitrarily-defined plasma given an initial gas density, composition, and incident spectrum. The code solves for the ionization, level populations, molecular state, thermal equilibrium, and emitted spectrum from the nebula. The source files and all necessary supporting data are available from \url{www.nublado.org}. Cloudy incorporates recombination coefficients from \cite{Badnell03} and \cite{Badnell06}, as well as ionic emission data from the CHIANTI collaboration database \citep[version 7.0, see][]{Dere97,Landi12}. 

In the following, we neglect dust and assume solar abundances for the gas phase metallicity, as given in Cloudy \citep[see ``Hazy'' documentation for the default values, taken from][]{GS98,AP01,Holweger01}. We assume a constant density in the gas and compute the gas temperature self-consistently. For a fixed source temperature, luminosity, and gas density, the size of any Str{\" o}mgren sphere is relatively insensitive to variations in the metallicity. We assume spherical symmetry in our calculations; for DEM L71 we carry out calculations assuming densities consistent with both the eastern and western rims, in order to account for the range in plausible Str{\" o}mgren radii. The age of the remnants considered here are all 1--2 orders of magnitude less than the local recombination timescale, therefore in the following we assume steady-state models.

The spectra of accreting, steadily nuclear-burning white dwarfs are well-approximated by blackbodies, except far into the Wien tail where carbon and oxygen edges have a pronounced effect in the soft X-ray band \citep{Rauch10}. Here we are concerned principally with the extreme UV emission (30\AA\ $\lesssim \lambda \lesssim $ 912 \AA), therefore we assume blackbody ionizing spectra in our models \citep[see discussion in][]{Chen15,WG16}.

White dwarfs accreting below the steady-burning threshold may also emit significant UV and soft X-ray emission, assuming accretion is mediated by a disk. Half of the gravitational binding energy released by infalling material must be emitted by the disk:

\begin{equation}
L = \frac{1}{2}\frac{G M_{\rm{WD}} \dot M}{R_{\rm{WD}}}
\end{equation}

\noindent where the white dwarf radius $R_{\rm{WD}}$ may be found as a function of white dwarf mass ($M_{\rm{WD}}$). Here we take the zero-temperature white dwarf models of \cite{Panei00}, as approximated by the fit given in \cite{GB10} and  \cite{Woods17}. For a Chandrasekhar-mass white dwarf, we take the radius of a 1.35$M_{\odot}$ model as an upper limit; smaller radii would produce larger luminosities. For the disk spectrum, we assume a Shakura-Sunyaev \citep{SS73} disk and find the spectral shape using the ezDiskBB model \citep{Zimmerman04} from the X-ray spectral modeling package XSPEC \citep{XSPEC}. 

Note that the disk luminosity considered in the following is independent of any additional high-energy flux associated with post-novae supersoft phases, and neglects the additional luminosity of the boundary layer \citep[see discussion in][]{Woods17}. Therefore, our estimates provide a lower bound on the extreme-UV/soft X-ray luminosity of accreting, non-steady-burning white dwarfs in the absence of significant obscuration from e.g., a disk wind (see discussion in section \ref{discussion}).

\subsection{Results}
\label{results}

Shown in Fig. \ref{allSNe} are the maximum progenitor luminosities, as a function of effective temperature, which would be consistent with observations of the conditions in the ISM surrounding the remnants SN 1006, 0509-67.5, 0519-69.0, and DEM L71. Also shown for reference are the steady hydrogen-burning white dwarf models of \cite{Wolf13}, as well as the temperatures and bolometric luminosities inferred for known close-binary supersoft X-ray sources with well-constrained luminosities and temperatures \citep{Greiner00,Starrfield04}. 

We find that the progenitor of SN 1006 can not have been simultaneously hot ($5\times 10^{4}K$ $\lesssim$ T $\lesssim 10^{6}$K) and luminous (L $\gtrsim 10^{36}$erg$\rm{s}^{-1}$) for $\tau _{\rm{rec}} \approx $ 100,000 years prior to explosion. At the highest temperatures, this is comparable to the deepest upper limit published for any recent, nearby supernovae based on the absence of a supersoft source in archival X-ray data, on the progenitor of SN 2011fe \citep{Nielsen12}; at temperatures comparable to observed supersoft sources, our upper limits lie 1--2 orders of magnitude below previous such X-ray constraints \citep[e.g.,][see Fig. \ref{allSNe} for comparison]{Nielsen13}. The high neutral helium fraction rules out known supersoft sources as well as theoretical models of steadily nuclear-burning accreting white dwarfs. Note that the earlier upturn in our helium constraint at low temperatures in Fig. 1 arises from the lower number of He-ionizing (E $>$ 24.6eV) and He II-ionizing (E $>$ 54.4eV) photons. 

\begin{figure}[t]
\centering
\includegraphics[width=0.5\textwidth]{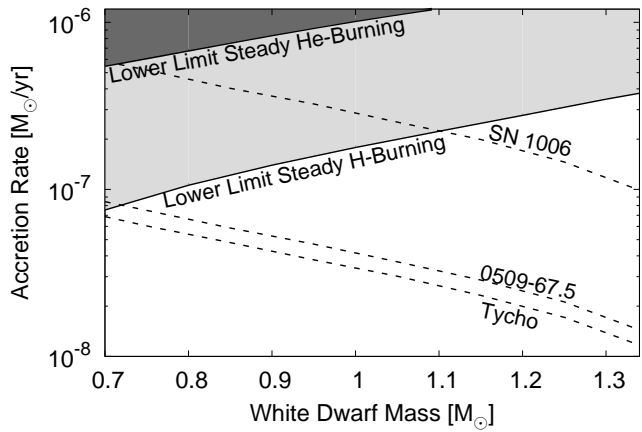}
\caption{Upper limits (dashed lines) on the accretion rate as a function of white dwarf mass for the progenitors of 0509-67.5 and SN 1006 given the ionization state of the surrounding ambient medium, assuming accretion mediated by an optically-thick, geometrically-thin (Shakura-Sunyaev) disk without additional nuclear-burning luminosity or significant obscuration (see discussion in text). For comparison, we have shown the same upper limit found for the progenitor of Tycho's supernova \citep{Woods17}, as well as the lower limits for steady nuclear-burning given hydrogen \cite{Wolf13} and helium \citep{Wang18} accretion. Light and and dark gray-shaded regions denote accretion rates for which the luminosity would be dominated by nuclear-burning on the white dwarf for H and He accretion, respectively.}\label{mdot}
\end{figure}

For non-nuclear-burning sources in which accretion is mediated via an unobscured disk, our models for photoionization by the disk's emission provide an upper limit on the allowable accretion rate for an approximately Chandrasekhar-mass white dwarf of $\dot M_{\rm{MAX}} \approx 9.5 \times 10^{-8} M_{\odot}/$yr (see also fig. \ref{mdot}). The high hydrogen ionization but low helium ionization is also incompatible with photoionization either by emission from the reverse shock or from the original supernova itself, and appears to be consistent with the ambient Galactic Lyman continuum \citep{Ghavamian02}. We cannot, however, exclude a recurrent nova binary with a relatively longer recurrence timescale (i.e., with $\dot M < \dot M_{\rm{MAX}}$), for which theoretical models suggest a white dwarf undergoing novae may still grow in mass \citep[e.g.,][]{Yaron05}.

The high neutral hydrogen fraction alone in the environment surrounding 0509-67.5 excludes any supersoft source progenitor in the $\sim10^{5}$ years prior to its explosion. Following the same argument as above, the maximum permissible accretion rate for a Chandrasekhar-mass white dwarf is $\dot M_{\rm{MAX}} \gtrsim 1.4\times10^{-8}M_{\odot}/$yr. This is comparable to the upper limit derived for the progenitor of Tycho's supernova \citep{Woods17}, and excludes any viable recurrent nova progenitor \citep{Yaron05}.

More generally, our upper limits on the accretion rate for the progenitors of 0509-67.5, SN 1006, and Tycho as a function of white dwarf mass are shown in Fig. \ref{mdot}. The gray shaded region denotes accretion rates above the lower-limit for steady hydrogen-burning; for stable accretion from e.g., a main sequence or red giant donor, here the hydrogen nuclear-burning luminosity dominates the emission (see discussion above). Note that for accretion rates $\gtrsim 3 \times$ the lower limit for steady-burning, the accreting white dwarf may enter the optically-thick wind regime \citep{HKN96}; as discussed above, this is excluded by the present evolutionary state of the remnants \citep{Badenes07}.

For stable helium accretion, the lower limit for steady nuclear-burning is nearly an order of magnitude greater \citep{PTY14,Brooks16,Wang18}. Helium-burning progenitors may also be excluded, as the lower energy release per unit mass of helium processed through nuclear-burning is offset by the higher accretion rates needed. We cannot exclude the slow accumulation of a helium shell at lower accretion rates \citep[as in, e.g., variations on ``double-detonation'' explosion models,][]{Nomoto82,Woosley86,Livne90}, however we constrain the maximum viable accretion rate, e.g., for Tycho and 0509-67.5, a $\gtrsim 1M_{\odot}$ progenitor could not have accreted at greater than $\sim 3-5 \times 10^{-8} M_{\odot}\rm{yr}^{-1}$.

For DEM L71 and 0519-69.0, the steady H-burning models of \cite{Wolf13} are excluded, although the luminosities and temperatures of known accreting nuclear-burning white dwarfs are observed in less luminous states. Therefore, we can not immediately exclude e.g., a low-luminosity or transient supersoft source as the progenitor of either DEM L71 or 0519-69.0. Deep limits on (or a measurement of) the [O III] 5007\AA\ flux, or similarly for any He I and He II recombination lines, would be able to confirm or exclude the existence of any such subluminous progenitor \citep{Ghavamian01,WG16}.

All upper limits on progenitor luminosities given above will depend on the measured shock radius, and the ionization state and density of the pre-shock ambient medium. Therefore, our results are sensitive to errors in the measured values of each. From eq. \ref{RS}, one may infer that at constant radius, any upper limit on the ionizing photon luminosity will scale with $\rm{n}_{0}^{2}$. For this reason, in each case we have chosen the highest pre-shock gas density consistent with the present-day size and evolutionary state of the remnant, rather than the best fitting value, in order to provide the most conservative constraint on the progenitor. Lower densities would yield even stronger upper limits.

Similarly, from eq. \ref{RS}, it is clear that the maximum ionizing photon luminosity for any progenitor will scale with the cube of the radius of the Balmer-dominated shock. Note however that this is independent of any asphericity in the observed remnant; in the ambient interstellar medium, the optical depth to e.g., 13.6eV photons is extremely high, and consequently any ionizing photons from ground-state recombinations are expected to be absorbed locally. This allows us to make the ``on-the-spot'' approximation \citep[Case B,][]{Osterbrock}, and consider only the attenuated flux of the source and any locally-produced photons in modelling the ionization state of the ISM at a radius r, in this case the measured radius of the Balmer-dominated shock (see section 2.2.1 and 2.2.2, and references therein).

\begin{figure}[t]
\centering
\includegraphics[width=0.5\textwidth]{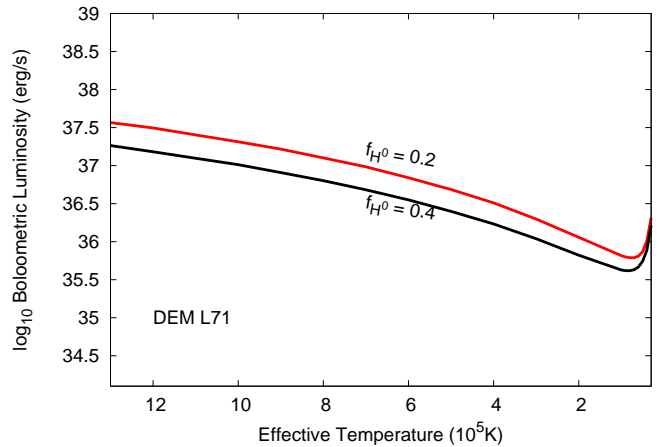}

\caption{Upper limit on the progenitor luminosity for DEM L71 as a function of temperature, for the maximum ($f_{H^{0}} = 0.4$) and minimum ($f_{H^{0}} = 0.2$) hydrogen neutral fractions consistent with observations of the western rim of the remnant. For all upper limits given in fig. \ref{allSNe}, the minimum neutral fraction/maximum ionized fraction is used.}\label{variation}
\end{figure}

In order to characterize the sensitivity of our upper limits to the measured ionization fraction \citep{Ghavamian02, Ghavamian03, Ghavamian07}, we have again plotted our constraint on the progenitor of DEM L71 (based on the western edge of the remnant), for the minimum (as in Fig. \ref{allSNe}) and maximum neutral fractions consistent with the work of \cite{Ghavamian07}. For effective temperatures $\gtrsim 4\times 10^{5}$K, our result scales linearly with the inferred neutral fraction, and only slightly more steeply at lower temperatures.
For this reason, throughout the text we take the minimum neutral fractions consistent with observations and numerical models of the Balmer-dominated shocks associated with each remnant (see $\S$2.2.1 \& $\S$2.2.2). This means the upper limits quoted above are the most conservative consistent with the observed optical features.

\subsection{Possible Caveats?}
\label{discussion}

In the preceding analysis, we have considered only ``unobscured'' sources, with photoionizing radiation being emitted either from the progenitor's surface or from a disk for $\sim \tau_{\rm{rec}}$ immediately prior to explosion. The question naturally arises whether any caveats to these assumptions may allow for an otherwise hot, luminous progenitor scenario such as the accretion channel to remain viable. In particular, could the ionizing emission be obscured near the source, or could a significant delay between the hot luminous phase and the explosion hide the companion? Here we address these issues in turn:

\begin{itemize}

\item Could the emission be obscured by a stellar wind?

If a white dwarf progenitor were accreting at a rate above the steady-burning regime \citep[e.g.,][]{Wolf13}, it may shed much of the accreting material in an optically-thick wind \citep{HKN96}; this could obscure the central source if the wind mass loss rate is sufficiently high. However, this would excavate a large (up to tens of parsecs) low-density cavity surrounding the progenitor, the existence of which may be excluded from hydrodynamical models of the remnants' evolution \citep{Badenes07}. In particular, the evolution of all remnants considered here is consistent with expansion into a uniform, undisturbed ISM \citep{Maggi16,PB17}. Similarly, a slow wind from a companion star  would require mass loss rates $\gtrsim 10^{-6} M_{\odot}\rm{yr}^{-1}$ in order to obscure a supersoft source's photoionizing emission \citep{NG15}. Such dense winds \citep[and indeed, giant companions,][]{Olling15} appear to be uncommon for Type Ia supernova progenitors, given the strong constraints on circumstellar interactions for other SNe Ia \citep[e.g.,][]{Chomiuk12,MMN14}. We conclude that obscuration of the progenitor's photoionizing emission by a dense wind from the progenitor or its companion does not appear to be viable for the remnants considered here.

\item Could the disk emission be obscured and/or reprocessed by an outflow from the disk itself?

If an accretion disk surrounding a massive white dwarf progenitor drives a dense wind, this may mask the disk's photoionizing emission, which would otherwise be expected even for modest accretion rates ($\gtrsim$ a few $\times 10^{-8}M_{\odot}\rm{yr}^{-1}$). This is the scenario proposed for M31N 2008-12a \citep{highmdot}, a recurrent nova in Andromeda with a recurrence time $\sim$ 0.5--1 year \citep{Darnley15,PeriodHenze15}. Comparing late-time photometry with numerical models for the accretion disk spectra, \cite{highmdot} infer a quiescent disk accretion rate of $\gtrsim 10^{-6} M_{\odot}\rm{yr}^{-1}$ -- greatly in excess of the white dwarf's accretion rate as inferred from modelling the novae themselves \citep[$\gtrsim 1.7\times 10^{-7}M_{\odot}\rm{yr}^{-1}$, see e.g.,][]{Tang14}. To account for this, \cite{highmdot} invoke a dense outflow from the disk, but do not model this or its effect on the emergent spectrum in detail. Further studies are necessary in order to reconcile the discrepancy between the inferred, high disk accretion rate for this system and theoretical models for the (much lower) threshold for steady nuclear-burning on white dwarf surfaces \citep[e.g.,][]{Wolf13}. We note, however, that at least three problems arise in invoking an M31N 2008-12a analog as the progenitor of any supernova remnant considered here: 
\begin{enumerate}
\item The post-nova supersoft phase associated with each eruption would produce a substantial time-averaged ionizing flux \citep[for M31N 2008-12a, $\Delta t_{\rm{SSS}} \sim 19$ days with a characteristic temperature $\sim$97eV,][]{XrayHenze15}. Depending on its precise recurrence time, spectral evolution, and long-term evolution, such a system may be excluded by the photoionization constraints provided above, e.g., a SSS phase duration of $\sim$10-15 days, peak luminosity of $\sim3\times 10^{37}$ erg$\rm{s}^{-1}$ and recurrence time of $\sim$12 months would be equivalent to a persistent source with luminosity of $\sim10^{36}$erg$\rm{s}^{-1}$, approximately the upper limit for a $10^{6}$K progenitor for SN 1006. Further modelling of the spectra and time evolution of this and other recurrent novae must be carried out before a more precise statement can be made.
 
\item If, as has been proposed for M31N 2008-12a \citep{novaremnant}, long-lived and frequently-erupting recurrent novae systems can excavate large ($\sim$100pc) cavities in their surrounding ISM, this is already excluded for all remnants considered here, both by their present evolutionary state \citep{Badenes07} and their present interaction with ISM of typical densities ($\sim 1\rm{cm}^{-3}$, see table \ref{data}).

\item As discussed above, radio and X-ray upper limits indicate SNe Ia do not interact with significant circumstellar material, such as from an outflowing disk wind \citep{Margutti14,Chomiuk16}. 
\end{enumerate}
\item Could there be a long delay between the hot luminous phase and explosion?

From the analysis presented here, we cannot exclude any ``spin-up/spin-down'' progenitor model \citep{Justham11}, in which an accreting white dwarf is spun-up by accretion until surpassing the Chandrasekhar limit, should the spin-down time until explosion exceed the recombination time. Note, however, that if most SNe Ia are preceded by a long spin-down time, this implies a large population of rapidly-spinning white dwarfs exists in the Milky Way. The dearth of such objects, as well as other issues with the spin-up/spin-down model, have previously been investigated extensively \citep[see e.g.,][]{DiStefano11,MMN14}, therefore we do not consider this model further here.

\end{itemize}

\vskip0.5cm
\section{Conclusions}
\label{conclusion}

Supernovae provide an invaluable probe of the ISM, even as they heat it and enrich it with heavy elements. Many SN Ia remnants are observed to produce Balmer-dominated shocks, as the advancing remnant overruns ambient neutral hydrogen. Modeling the broad-to-narrow flux ratios of hydrogen Balmer lines in these shocks, as well as measurements of other emission lines either in the shocked gas or in diffuse emission ahead of it, can strongly constrain the ionization state of the gas in the vicinity of these explosions \citep[e.g.,][]{Ghavamian00}. This is particularly true of measurements of He I and He II emission, which can strongly constrain the helium ionization fraction in the surrounding ambient medium \citep{Ghavamian02}.

Here, we have demonstrated that such measurements can provide a powerful diagnostic of the nature of the progenitors of SNe Ia, constraining their luminosities and temperatures for the last $\approx$ 100,000 years. A close binary supersoft source is excluded as the progenitor of SN 1006 by the high neutral helium fraction in the surrounding ISM. The low density and ionization state of hydrogen in the vicinity of 0509-67.5 excludes a supersoft source as well as any accreting white dwarf with $\dot M_{\rm{MAX}} \gtrsim 1.4\times10^{-8}M_{\odot}/$yr.  The environments surrounding DEM L71 and 0519-69.0 are, however, only marginally in tension with known supersoft sources, and a recurrent nova progenitor remains plausible. Future deep narrow-band observations centred on [O III] 5007\AA, [O I] 6300\AA, He I 6678\AA, and He II 4686\AA\ will be able to conclusively measure or rule out the presence of any relic nebula ionized by the progenitor of these and many other nearby supernovae.

\acknowledgments

The authors would like to thank the gracious hospitality of the organizers of the conference ``Supernova Remnants: An Odyssey in Space After Stellar Death,'' during which this work began, and the Lorentz centre and all attendees of the workshop "Observational Signatures of Type Ia Supernova Progenitors III," for many fruitful discussions, as well as the anonymous referee for helpful comments which improved the manuscript. The work of P. G. was supported by grants HST-GO-12545.08 and HST-GO-14359.011. C. B. acknowledges support from grants NASA ADAP NNX15AM03G S01 and NSF/AST-1412980. M. G. acknowledges partial support by Russian Scientific Foundation (RNF) project 14-22-00271.

\bibliographystyle{apj}

\end{document}